\documentclass[aps,prb,showpacs,reprint,amsmath,amssymb,amsfonts,superscriptaddress]{revtex4-1}
\usepackage{graphicx}
\usepackage[T1]{fontenc}
\usepackage{hyperref}
\newcommand{\LSMO}{La$_{1-x}$Sr$_{1+x}$MnO$_4$}
\newcommand{\LSMOhd}{La$_{0.5}$Sr$_{1.5}$MnO$_4$}

\begin{document}

\title{Signatures of electronic polarons in {La$_{1-x}$Sr$_{1+x}$MnO$_4$} observed \\ by electron energy-loss spectroscopy}

\author{Roberto Kraus}
\author{Matthias Schrade}
\author{Roman Schuster}
\author{Martin Knupfer}
\affiliation{IFW Dresden, P.O. Box 270116, D-01171 Dresden, Germany}
\author{Alexandre Revcolevschi}
\affiliation{Laboratoire de Physico-Chimie de l'\'Etat Solide, Universit\'e Paris-Sud, F-91405 Orsay Cedex, France}
\author{Bernd B\"uchner}
\author{Jochen Geck}
\affiliation{IFW Dresden, P.O. Box 270116, D-01171 Dresden, Germany}

\date{\today}

\pacs{71.38.--k, 75.25.Dk, 79.20.Uv}

\begin{abstract}
The dielectric properties of  {La$_{1-x}$Sr$_{1+x}$MnO$_4$} single crystals with $x=$0, 0.125, 0.25, and 0.5 were studied by means of electron energy-loss spectroscopy as a function of temperature and momentum transfer. A clear signature of the doped holes is observed around 1.65\,eV energy loss, where spectral weight emerges with increasing $x$. For all $x\neq 0$,  this doping-induced excitation can propagate within the $ab$-plane, as revealed by a clear upward dispersion of the corresponding loss peak with increasing momentum transfer. The hole-induced excitation also shifts to higher energies with the onset of magnetic correlations for $x=0.5$, implying a strong coupling of charge and spin dynamics.
We conclude that (i) the loss feature at 1.65\,eV is a signature of electronic polarons, which are created around doped holes, and that (ii) this low-energy excitation involves the charge transfer between manganese and oxygen. The finite dispersion of these excitations further indicates significant polaron-polaron interactions.
\end{abstract}

\maketitle

\section{Introduction}

The doped manganites are well-known for their complex electronic behavior, encompassing the famous colossal magneto resistance,\cite{Helmolt1993,Jin1994,Salamon2001} electronic order \cite{Moritomo1995,Mizokawa1996} and unusual charge dynamics.\cite{Mannella2007}
These phenomena are due to strong interactions between lattice, charge, spin and orbital degrees of freedom, which often compete and result in unconventional electronic behavior.\cite{Tokura2000}
A typical consequence of such a competition is that a small external perturbation, corresponding to merely 1\,meV, can knock over the delicate balance between different interactions of the order of 1\,eV and, for instance, trigger an insulator-metal transition.
This entanglement of low and elevated energy scales was clearly observed in recent resonant inelastic x-ray scattering experiments.\cite{Grenier2005,Weber2010c} There it was found that the magnetic ordering in doped manganites, which is typically associated to 1\,meV physics, has strong effects on the electronic excitations up to about 24\,eV.
In this situation complex electronic behavior can arise, which can not be captured by well-established concepts used to describe conventional metals and insulators. It therefore becomes important to study the electronic system on low energy scales around 1\,meV as well as high energy scales around 1\,eV.

In this paper we address the charge dynamics of single layered \LSMO\/  at energies of the order of 1\,eV  by means of electron energy-loss spectroscopy (EELS). \LSMO\/ was chosen as a model system, because its well defined lattice structure and reduced dimensionality significantly simplify the analysis of the data and the development of theoretical models. The undoped parent compound of this family, LaSrMnO$_4$ ($x=0$), exhibits a G-type antiferromagnetic insulating low-temperature phase.\cite{Larochelle2005} The dielectric properties of this material were investigated by optical ellipsometry and the data were described in terms of an effective model.\cite{Goessling2008} Within this model for the undoped parent compound the effective Mott-Hubbard (MH) excitations were found at lower energies than the charge transfer (CT) excitations. It is important to point out that this assignment does not imply that the hybridization between oxygen 2p (O:2p) and manganese 3d (Mn:3d) states plays no role. In fact, the corresponding effective Hubbard model was introduced for the case of $U\sim \Delta$ and strong hybridization.\cite{Goessling2008}

The G-type antiferromagnetic insulating ground state of LaSrMnO$_4$ is quickly suppressed upon doping and around $x=0.5$ a complex electronically ordered phase, the so-called CE-phase, is observed at low temperatures.\cite{Moritomo1995,Murakami1998,Sternlieb1996} Even though this CE-phase is often discussed in terms of spin, charge and orbital order, it should be mentioned that the full charge disproportionation into  $\mathrm{Mn}^{3+}/\mathrm{Mn}^{4+}$ sites is not realized in the real material. Instead, a number of studies clearly indicate that the Mn-valence differences are small or even completely absent.\cite{Ferrari2003,Garcia2001,Luo2007,Herrero-Martin2010} 

Neutron scattering studies of \LSMO\/ with $x=0.5$ further showed that magnetic short range correlations develop in a complex fashion upon cooling and evolve into the CE-ordered phase with decreasing temperature.\cite{Senff2008} The presence of  electronic short range correlations is expected to have a strong impact on the charge carrier dynamics. Indeed, inelastic neutron scattering studies on a related bilayer material (La,Sr)$_3$Mn$_2$O$_7$ close to half doping showed  that dynamic CE-type correlations are present even in the metallic phase of this material \cite{Weber2009} and might be related to the unconventional charge dynamics probed by angular resolved photoemission spectroscopy.\cite{Mannella2007}

Unlike most of the other manganite materials, \LSMO\/  never becomes metallic for $0 \leq x \leq 0.6$, underlining the relevance of electronic correlations in these compounds. Furthermore, X-ray absorption studies (XAS) on the same doped \LSMO\/ crystals studied here, showed very clearly that hole doping not only creates unoccupied O:2p states, i.e., holes on oxygen, but also causes a major change of the orbital occupation at the manganese sites.\cite{Merz2006} These observations demonstrate the relevance of Mn:3d--O:2p hybridization and show that the doped holes have a strong O:2p character. The latter observation is in line with small or absent Mn-valence modulations described above.

In this paper we present a doping dependent EELS study of \LSMO , showing that a new peak emerges in the measured loss function upon increasing $x$. Different from what was observed in resonant inelastic scattering experiments on doped manganites,\cite{Grenier2005} the excitation observed here only exists in the doped samples and hence can be directly associated to the doped charge carriers. Based on our observations we conclude that the doping induced excitation observed by EELS is a signature of electronic polarons created upon hole doping.

This paper is organized as follows: In Sec.\,II we provide some experimental details regarding the EELS measurements. Sec.\,III and IV are then devoted to a description of the obtained results and to the discussion, respectively. Our conclusions are summarized in Sec.\,V.

\section{Experimental}\label{exp}

We investigated the dielectric properties of \LSMO\/ (LSMO) single crystals with $x=0$, 0.125, 0.25 and 0.5 in the temperature range between room temperature and 30\,K by means of EELS. This method enables to measure the loss function
\begin{equation}
\mathcal{L}(\mathbf{q},\omega)=\Im \left(-\frac{1}{\varepsilon(\mathbf{q},\omega)}\right)=
\frac{\varepsilon_2(\mathbf{q},\omega)}{\varepsilon_1^2(\mathbf{q},\omega)+\varepsilon_2^2(\mathbf{q},\omega)} \label{eqn1}
\end{equation}
where $\omega$,  $\mathbf{q}$ and $\varepsilon(\mathbf{q},\omega) = \varepsilon_1(\mathbf{q},\omega) + \mathrm{i} \varepsilon_2(\mathbf{q},\omega)$ denote the energy loss, the momentum transfer and the complex dielectric function, respectively.
Therefore a direct comparison between EELS spectra and the optical conductivity $\sigma\propto\omega \epsilon_2$ is not possible,  i.e., the same excitation leads to different peak positions in EELS and in the optical conductivity.
This can be easily understood in terms of the Drude-Lorentz model for a single oscillator centered at $\omega_0$ with a strength represented by $\omega_p$. In the loss function, this oscillator will result in a peak centered at $\omega = \sqrt{\omega_p^2+\omega_0^2}$, whereas in the optical conductivity it will cause a peak at $\omega = \omega_0$.  In general, as compared to optical conductivity, the corresponding peak in the loss function is shifted to higher energies, and this effect becomes more and more significant the lower excitation energy is. Another difference between EELS and optics is that EELS enables to perform momentum dependent measurements to access, for instance, optically forbidden excitations \cite{Atzkern2000,Knupfer2000} or the dispersion \cite{Schuster2007,Neudert1998} of excitations.

Single crystals of La$_{1-x}$Sr$_{1+x}$MnO$_4$ have been grown using the floating zone method.\cite{Reutler2003} For the EELS studies in transmission, thin and single crystalline films of approximately 100\,nm thickness are necessary. These films were cut from a single crystal using an ultramicrotome and mounted onto standard transmission electron microscopy grids. The orientation and quality of the samples was verified {\it in situ} by elastic electron diffraction.

The EELS measurements in transmission were performed with a dedicated spectrometer, which is described in Ref.\,\onlinecite{Fink1989}. The primary beam energy was 172\,keV, the energy and momentum resolution were set to 90\,meV and 0.04 \,\AA$^{-1}$, respectively. Note that at such a high primary beam energy only singlet excitations are possible.\cite{Fink1989} The spectrometer is equipped with a flow-cryostat, allowing for temperature dependent measurements down to 15\,K.

\begin{figure}[t!]
\includegraphics[width=0.40\textwidth]{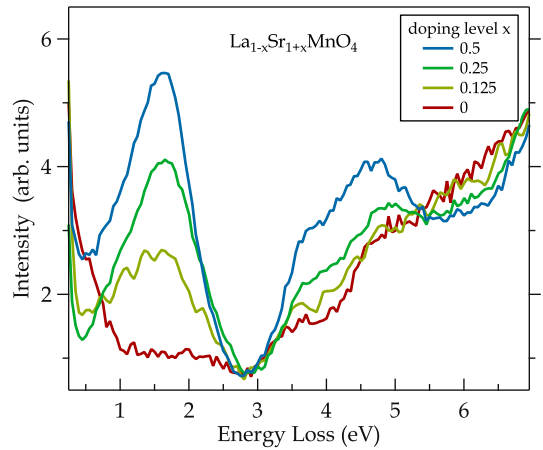}
\caption{(Color online) EELS intensity of La$_{1-x}$Sr$_{1+x}$MnO$_4$ in the optical limit at $q=0.1$\, \AA$^{-1}$ for doping levels $0 < x < 0.5$. The measurements were done at room temperature. The spectra are normalized to the volume plasmon excitation at 14\,eV.}
\label{fig_N1}
\end{figure}

\section{Results}

The doping dependent EELS intensity of LSMO in the optical limit at $q=|\mathbf{q}|=0.1$\,\AA$^{-1}$ is presented in Fig.\,\ref{fig_N1}. It can clearly be observed that a prominent new excitation appears at 1.65\,eV with increasing hole doping $x$. Also at higher energy losses between 3 to 5.5\,eV the spectral weight is increased with doping. The observed integrated spectral weight in both energy regions increases roughly linearly with $x$. Furthermore no anisotropy in the spectra were found within the $ab$-plane as expected for a tetragonal symmetry.

\begin{figure*}[t!]
\includegraphics[height=0.24\textheight]{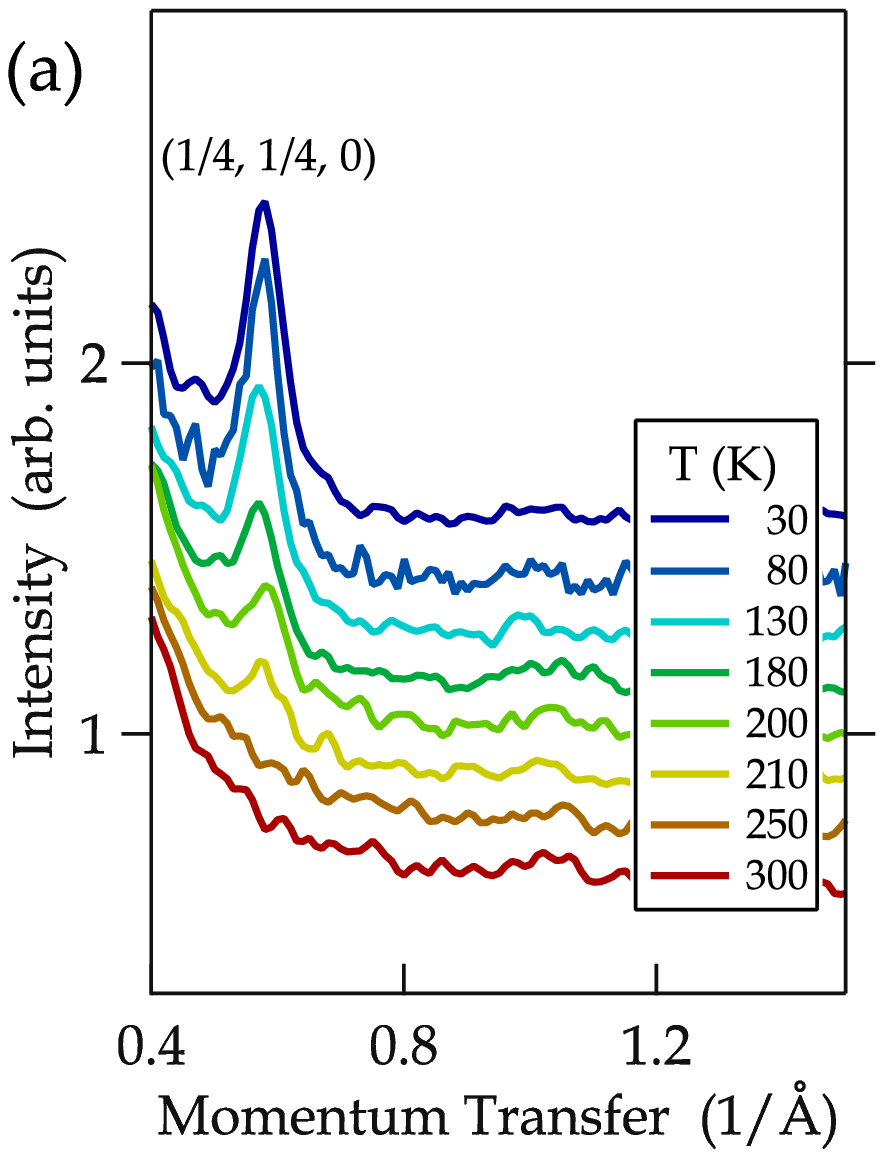}
\includegraphics[height=0.24\textheight]{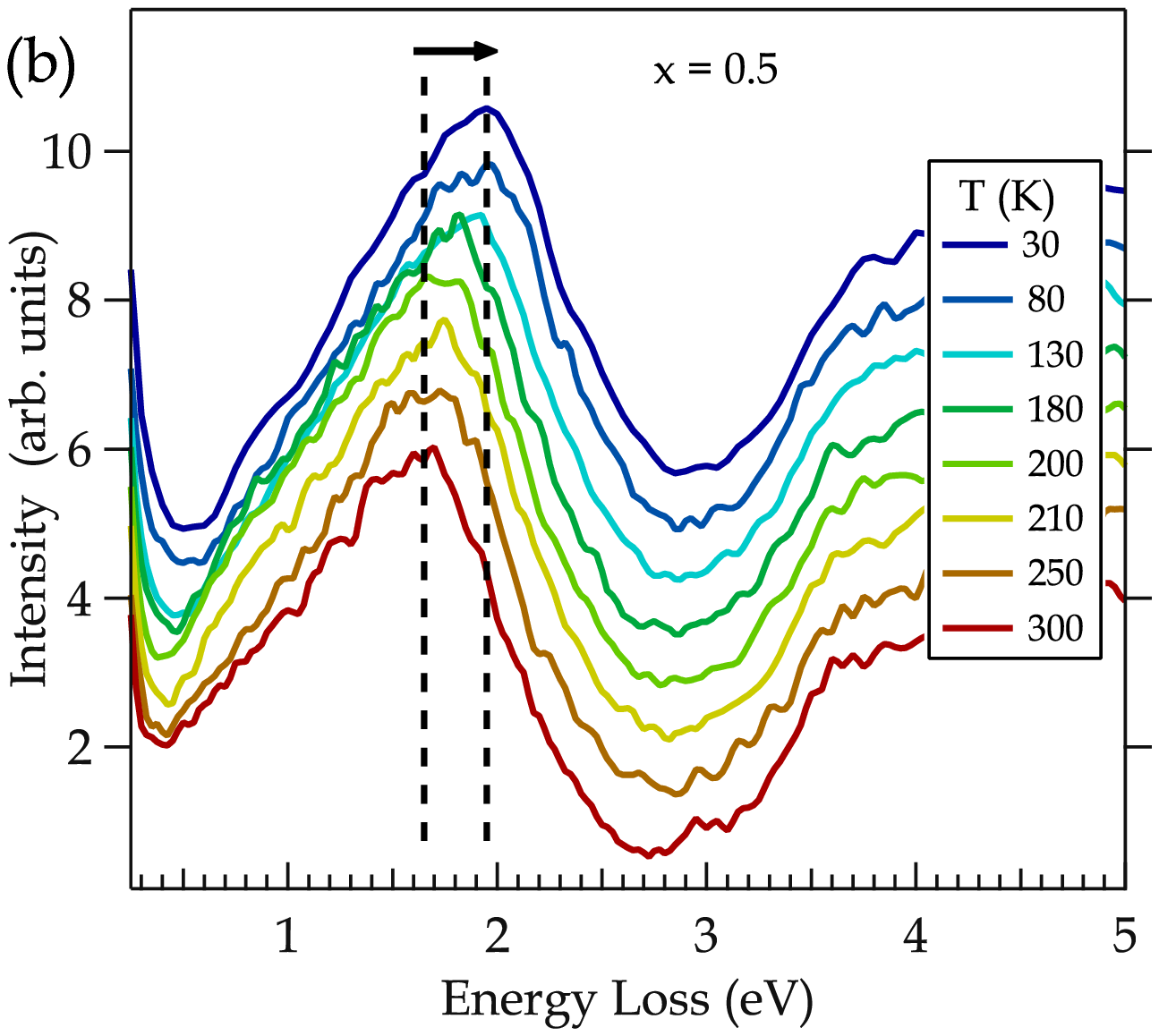}
\includegraphics[height=0.24\textheight]{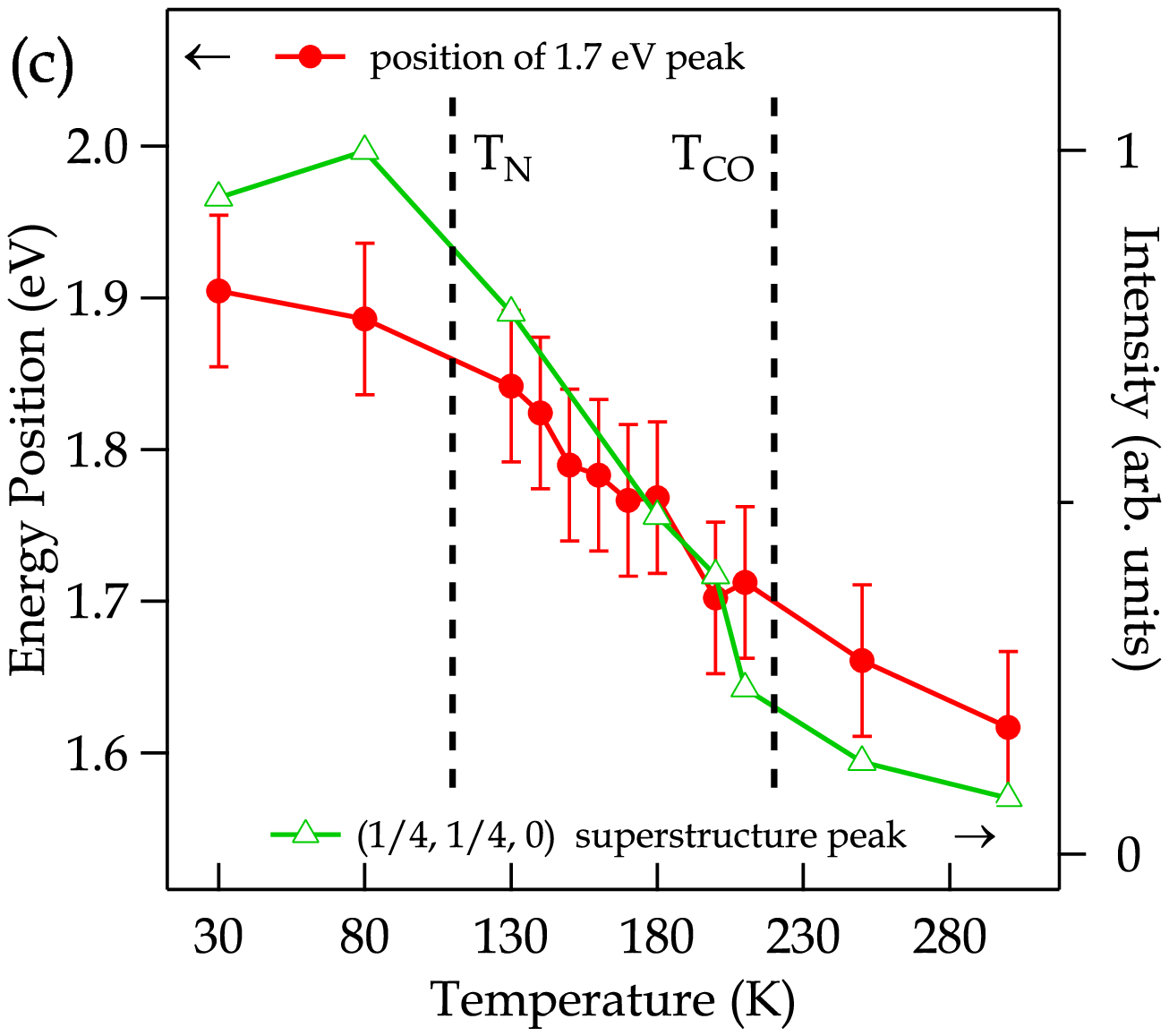}
\caption{(Color online) Temperature dependent EELS and electron diffraction data for the half doped material 
\LSMOhd. (a) Electron diffraction measurements. Upon entering the electronically ordered phase, the orbital order superstructure peak develops at $(\frac{1}{4} ,\, \frac{1}{4} ,\, 0)$ below 210\,K .
(b) EELS spectra in the optical limit taken at different temperature. The first excitation shifts by $\approx 300$\,meV to higher excitation energies. The spectra have been  shifted vertically for clarity.
(c) Comparison of the energy position of the first loss peak and the intensity of the superstructure peak. The intensity of the peak is determined by the area under the superstructure  peak.}
\label{fig_phase}
\end{figure*}

As already mentioned in the introduction, half-doped \LSMOhd\/ exhibits an electronically ordered groundstate at low temperatures. Specifically, upon cooling orbital order is  established first at 225\,K, while the spin order becomes long ranged only below 110\,K.\cite{Merz2006,Murakami1998}  In Fig. \ref{fig_phase}\,a) the phase transition into the orbitally ordered phase is signaled by the (1/4,1/4,0) superlattice reflection, which was measured {\it in situ} by electron diffraction. The fact that this orbital order peak occurs around the bulk transition temperature shows that the film sample exhibits bulk-like properties. We could not observe the (1/2,1/2,0) reflection, which is known to occur at the same temperature as the (1/4,1/4,0) peak.
However, zone-axis electron diffraction experiments show that the (1/2,1/2,0) is significantly weaker than the (1/4,1/4,0) superlattice reflection.\cite{Moritomo1995}
From this we conclude that the intensity of the (1/2,1/2,0) is too low to be detected in our {\it in situ} electron diffraction experiment, which is not optimized for the detection of weak reflections.
We note here in passing that all studied samples displayed similar quality with respect to the crystallinity.

Upon cooling down from room temperature to 30\,K significant changes were  observed in the loss function of \LSMOhd\/ as shown in Fig.\,\ref{fig_phase}\,b). As can be observed in the figure, the peak in the loss feature at 1.65\,eV shifts by about 300\,meV to higher energy losses upon cooling. The comparison of the  intensity at (1/4,1/4,0) and the position of the doping induced loss peak shown in Fig.\,\ref{fig_phase}\,c) indicates a direct relation of these two quantities and reveals that the most pronounced changes of the loss function occur in the temperature range between 220\,K and 120\,K, i.e., exactly in the temperature range where magnetic correlations develop.\cite{Senff2008}  
In addition, this temperature dependence is unique for $x=0.5$. For all the other investigated doping levels $x\neq0$ we observe only small changes of the loss function with temperature and shifts of  at most of 100\,meV for the 1.65\,eV peak. Since $x$=0.125 and 0.25 do not display long-range electronic order, the temperature dependence observed for $x=0.5$ must be related to the formation of the long range ordered CE-phase below 110\,K.

\begin{figure}[t!]
\begin{center}
\includegraphics[width=0.4\textwidth]{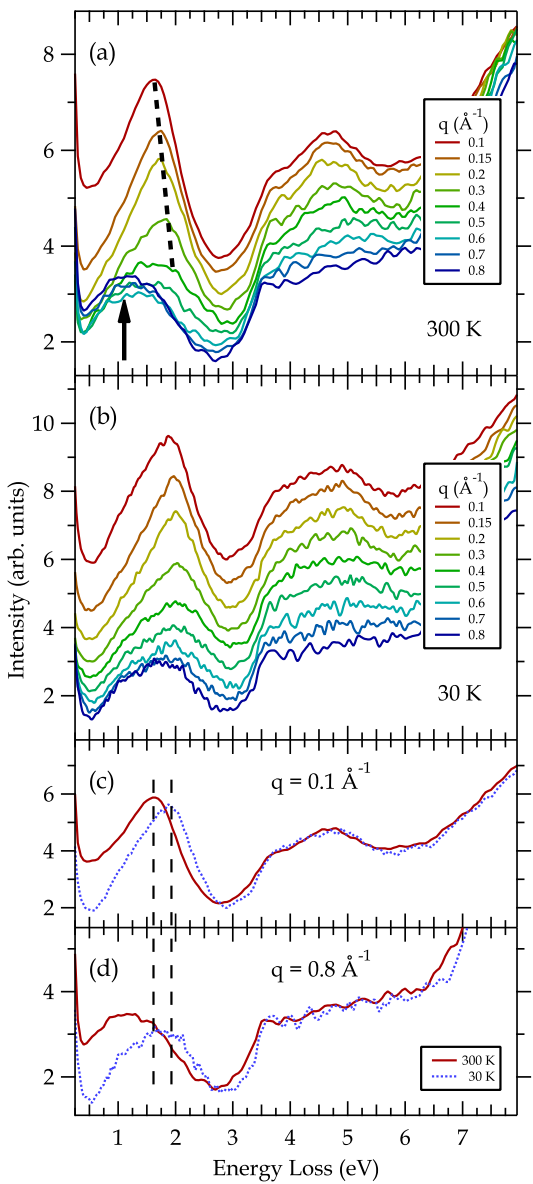}
\caption{(Color online) Momentum dependence of the EELS intensity for $x=0.5$ measured at 300\,K (a) and 30\,K (b). The momentum transfer is parallel to the $\Gamma$X-direction of the tetragonal Brillouin zone. For the room temperature measurement the quadrupole excitation at 1.2\,eV is indicated by the arrow and the dispersion of 200\,meV from the first peak at 1.65\,eV is highlighted by the dashed line in the top panel. In the two bottom panels the loss functions at low (c) and high (d) momentum transfer are compared  for 300\,K and 30\,K.  The energy position of the first excitation is highlighted at both temperatures.}
\label{fig_disp_temp}
\end{center}
\end{figure}

In Fig.\,\ref{fig_disp_temp} momentum dependent measurements are presented for the half doped sample at room temperature and in the ordered state at 30\,K. We will start with the description of the room temperature data. As can be seen in Fig.\,\ref{fig_disp_temp}\,a), the loss feature around 1.65\,eV exhibits a discernible upward dispersion by about 200\,meV with increasing $q$, which shows that the corresponding excitations can propagate within the $ab$-plane. The width of the loss peak at 1.65\,eV together with the strong reduction of the spectral weight with increasing $q$, however, hinder a detailed determination of the dispersion. The decreasing intensity of the 1.65\,eV feature identifies it as a dipole allowed transition.\cite{Knupfer1999a} Remarkably, an additional excitation at lower energies is found around 1.25\,eV at large $q$. This intensity increase with $q$ is typical for higher order transitions, like a quadrupole transition for example, which are forbidden in the optical limit and have an intensity  maximum at higher momentum transfers.\cite{Knupfer1999a} The dispersion behavior described above for $x=0.5$ at room temperature was found to be universal for all doping levels.

In the electronically ordered phase of \LSMOhd\/  (Fig.\,\ref{fig_disp_temp}\,b), the dispersion of the dipole part up to $q\simeq0.4$\,\AA$^{-1}$ is very similar to that obtained at room temperature, apart from the overall upshift of about 300\,meV described earlier (Fig.\,\ref{fig_phase}\,b).
There are however clear temperature dependent changes of the spectral features at large $q$. As demonstrated in  Fig.\,\ref{fig_disp_temp}\,c)-d), the intensity of the multipole peak at higher $q$ is reduced at low T and possibly shifted by more than 300\,meV to higher energies than the rest of the spectrum. Unfortunately the broad structure of the peak again prevents a more detailed analysis of this modification. The data given in Fig.\,\ref{fig_disp_temp}\,c)-d) also shows that the temperature dependent changes of the $q$-dependent loss function are completely confined to the low-energy region. The spectral weight above 3\,eV remains essentially unchanged.

\section{Discussion}

As described above, the major change in the loss function with doping is the appearance of a new excitation at 1.65\,eV. Excitations in this energy range were also observed by resonant inelastic x-ray scattering (RIXS) at the Mn K-edge.\cite{Grenier2005,Inami2003,Ishii2004,Weber2010c} Quite similar to the behavior observed in EELS, the RIXS excitations  are also affected by magnetic order and show some dispersion. There is, however, one clear difference between RIXS and EELS: in contrast to EELS, the RIXS spectra show a strong excitation around 2\,eV for {\it all} doping levels, including $x=0$. This shows that RIXS and EELS provide quite different cross sections for different types of excitations. This becomes particularly clear by comparing spectra for the undoped compound. Using RIXS a strong excitation is observed at 2\,eV for $x=0$,\cite{Inami2003} whereas the EELS data displayed in Fig.\,\ref{fig_N1} only exhibits very weak features.

This conjecture is strongly supported by the assignment of the EELS features for $x=0$, which is obtained by a comparison to optical ellipsometry: calculating $\mathcal{L}$ from the $\epsilon$ measured by ellipsometry\,\cite{Goessling2008} (cf. Eqn.\,\ref{eqn1}), we found excellent agreement with experimental EELS data measured in the optical limit ($q\rightarrow0$). This allows to map the types of excitations identified in Ref.\,\onlinecite{Goessling2008} onto our EELS spectrum. According to this mapping, the weak EELS-features at energy-losses between 1.5 and 2\,eV of the undoped sample are due to effective MH excitations between $3z^2-r^2$ and $x^2-y^2$ orbitals on neighboring sites. Intersite dd-excitations of this type would indeed be strongly enhanced in hard x-ray RIXS, due to the screening of the core-hole in the intermediate state.\cite{Kotani2001,Brink2006,Haverkort2010} For EELS and optics, where no such intermediate state is involved, the cross section is much smaller, as observed experimentally. RIXS and EELS therefore probe different types of excitations with different cross sections. This complementarity of RIXS, on the one hand, and EELS (and optics), on the other hand, is important, as it enables to disentangle excitations overlapping in energy.

In the present case, the weak sensitivity of EELS to intersite dd-excitations allows to clearly observe the doping induced excitations below 3\,eV. This doping dependence shown in Fig.\,\ref{fig_N1} reveals that the spectral weight peaked at 1.65\,eV is directly related to the doped holes. In optical studies on other doped manganites, corresponding excitations were observed around 1.1\,eV.\cite{Lee2007a,Jung2000a,Woodward2004} These observations indicate that the doping induced excitations observed here for \LSMO\/ are a generic feature of doped manganites.

In most of the previous studies, the excitations below 3\,eV energy loss were discussed in terms of intersite $d^4d^4 \rightarrow d^3d^5$ transitions.\cite{Lee2007a} Even though this interpretation seems to be correct for the 2\,eV excitation observed in Mn K-edge RIXS, it cannot explain the doping dependence found in the present EELS study. A straightforward candidate would be an intersite excitation of the type $d^3d^4\rightarrow d^4d^3$. However, neglecting the interactions between neighboring MnO$_6$ octahedra, an upper bound for the resulting optical gap would be $\Delta_{e_g}/2\simeq 0.3$\,eV. This is much lower than the experimental value of 1.1\,eV.\cite{Lee2007a,Jung2000a} Another possible reason for the doping induced spectral weight are local charge transfer excitations: $d^3\rightarrow d^4\underline{L}$. But these excitations should not yield a strong polarization dependence, which again disagrees with the experimental data.\cite{Jung2000a,Woodward2004}

\begin{center}
\begin{figure}[t!]
\includegraphics[width=0.36\textwidth]{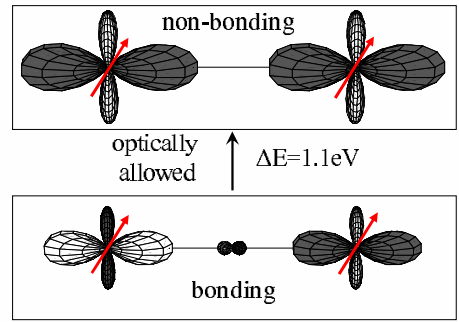}
\caption{(Color online)  Optical excitation of a Zener Polaron, consisting of 2 Mn-sites with ferromagnetically aligned spins and one O-site in between (further explanations in the text). The size of the orbital lobes represents the hole occupation at a given site. An optically allowed transition between the bonding and the non-bonding state is obtained at 1.1\,eV. This transition results in the 1.1\,eV peak in the optical conductivity, which corresponds to the peak at 1.65\,eV n in the loss function (see discussion in the text and in the introduction).}
\label{fig_pol}
\end{figure}
\end{center}

The experimental facts can, however, naturally be explained in terms of strong hybridization and polarons, which are formed around a doped hole. To illustrate this, we discuss the specific case of a so-called Zener Polaron (ZP).\cite{Zener1951} A ZP can be regarded as a bonding state that is formed by a doped oxygen hole and the two neighboring manganese sites (cf. Fig.\,\ref{fig_pol}). 
We consider a simplified configuration interaction model for an isolated purely ferromagnetic polaron with fixed orbital occupation at the Mn-sites. The following configurations of this Mn$_\mathrm{A}$--O--Mn$_\mathrm{B}$ cluster are taken into account:
$|1\rangle=|d_A^3\,p^2\,d_B^4\rangle$ (hole on Mn$_\mathrm{A}$), $|2\rangle=|d_A^4\,p^1\,d_B^4\rangle$ (hole on oxygen), and $|3\rangle=|d_A^4\,p^2\,d_B^3\rangle$ (hole on Mn$_\mathrm{B}$). It is further assumed that only the $e_g$-electrons can hop, while those in the $t_{2g}$ levels are completely localized. The hybridization between the O:2p and Mn:$e_g$ states is described in terms of a tight binding approximation,\cite{Slater1954} yielding the following Hamiltonian for the hopping of the doped hole:
\begin{equation}
\mathcal{H}=
\begin{pmatrix}
0 & pd\sigma & 0 \\
pd\sigma & \Delta & -pd\sigma \\
0 & -pd\sigma & 0\\
\end{pmatrix}
\end{equation}

In the above expression, $pd\sigma$ describes the hoping between the $|3x^2-r^3\rangle$ and the $|p_x\rangle$ state and $\Delta$ is the charge transfer gap. Using $pd\sigma$=-1.6\,eV and $\Delta$=3.5\,eV, which are close to what can be found in the literature,\cite{Bocquet1996} we immediately get an optically allowed transition at $\Delta E=\left(\sqrt{\Delta^2+8\,pd\sigma^2}-\Delta\right)/2$=1.1\,eV (cf. Fig.\,\ref{fig_pol}). 

Despite its simplicity, the above model provides a remarkable agreement with experiment: (i) The obtained optically allowed transition will result in a peak of the optical conductivity at 1.1\,eV, as observed in experiment. \cite{Lee2007a}  As discussed above, the peak in the optical conductivity at 1.1\,eV  corresponds to the loss feature at 1.65\,eV observed by EELS (cf. Eqn.\,\ref{eqn1}). In other words, the model agrees well with EELS and optics. (ii) The model naturally explains the occurrence of a doping induced transition in optics and EELS. (iii) The model agrees perfectly with the polarization dependence found in optical experiments, \cite{Jung2000a,Woodward2004} because this transition is only allowed, if the projection of the polarization onto the Mn--O--Mn bond is non-zero. (iv) The model yields a strong O:2p character of the doped holes, as observed in XAS.\cite{Merz2006} (v) It is consistent with the doping induced change in the orbital occupation at Mn, which was also observed by XAS.\cite{Merz2006} (vi) The obtained strong O:2p character of the doped holes provides a natural explanation for the small changes of the Mn-valence as a function of temperature and doping found for the Mn-sites.\cite{Herrero-Martin2010}

The above model is able to reconcile a number of experimental observations. It is clear, however, that this local description constitutes a simplification. A more realistic description would need to account for a system of interacting polaronic charge carriers on a periodic lattice, including the orbital degrees of freedom.
In fact, a previous theoretical study used such an approach to describe the optical properties of half-doped manganites.\cite{Cuoco2002} Topological frustration in the CE-phase results in a situation where the electronic system can be broken down into interacting polaronic objects.\cite{Brink1999} This decomposition led to a realistic description of the dielectric properties and showed that the excitations of the polarons can propagate as a result of polaron-polaron interactions.\cite{Cuoco2002} In the present study a dispersion of the excited states is indeed observed. However, while this theoretical model refers to the long range ordered CE-phase, we observe the corresponding features and dispersions also in disordered phases. This indicates that interacting polaronic charge carriers are present in the materials without long range electronic order.

In many theoretical studies, including the one just described, the oxygen states are not taken into account explicitly. The available experimental data together with the simplified model described here, however, indicate that the O:2p character of the doped holes is important to understand the charge response of the doped manganites. This motivates further theoretical efforts that take the O:2p character explicitly into account.

We also note that the observed excitation energy and polarization dependencies alone do not allow to determine the detailed microscopic structure of the polarons in detail. For instance, instead of being located on a single Mn--O--Mn bond, the hole could occupy a $x^2-y^2$-symmetry state involving the $p$-states of all four in-plane oxygen sites next to a Mn-site. Such a polaron model would yield a similar agreement with experiment.
Based on the temperature dependence of the loss function at large momentum (Fig.\,\ref{fig_disp_temp}), one may even speculate that the microscopic structure of the electronic polarons depends on temperature, doping and maybe also on the specific manganites material. 

Nonetheless, according to the above discussion, the doping-induced excitations seen in EELS can be well understood in terms of charge-transfer type excitations, which are determined by the charge transfer gap $\Delta$ and the $pd$-hybridization $pd\sigma$. Complementary to this, Mn K-edge RIXS also probes Mott-type intersite $dd$-excitations at slightly higher energies, as discussed above.

\section{Conclusions}

To conclude, we have reported a strong doping dependence of the loss function around 1.65\,eV. The comparison between RIXS and EELS revealed that the two methods exhibit different cross sections to different excitations that occur in the same energy range between 1\,eV and 2.5\,eV energy loss. RIXS and EELS are therefore complementary techniques and provide a unique opportunity to disentangle the complex excitation spectrum of doped manganites. Based on the data presented here, previous studies, and a simplified configuration interaction model, we further conclude that the loss function shows clear signatures of electronic polarons that are created upon hole doping. We also observe a finite dispersion of the corresponding energy loss feature, implying that the excitations of these polarons can propagate within the $ab$-plane of \LSMO. Our analysis further indicates that the low-energy EELS-excitation at 1.65\,eV involves the charge transfer between oxygen and manganese. Accordingly, these charge transfer type excitations play an important role for the collective charge dynamics of doped manganites in this energy region.  Our results further support the notion that interacting electronic polarons are a generic feature of doped manganites.

\begin{acknowledgments}
The authors would like to thank George Sawatzky for very valuable discussions. We thank R. H\"ubel, S. Leger and R. Sch\"onfelder for technical assistance. JG gratefully acknowledges the support by the DFG through the Emmy-Noether program. RS acknowledges the support by the DFG through project KN393/13. 
\end{acknowledgments}

%

\end{document}